# Rank analysis of most cited publications, a new approach for research assessments

**Alonso Rodríguez-Navarro**[a,b,*] **and Ricardo Brito**[b]

[a]Departamento de Biotecnología-Biología Vegetal, Universidad Politécnica de Madrid, Avenida Puerta de Hierro 2, 28040, Madrid, Spain

[b]Departamento de Estructura de la Materia, Física Térmica y Electrónica y GISC, Universidad Complutense de Madrid, Plaza de las Ciencias 3, 28040, Madrid, Spain

*Corresponding author. E-mail address: alonso.rodriguez@upm.es

**ABSTRACT**

Citation metrics are the best tools for research assessments. However, current metrics may be misleading in research systems that pursue simultaneously different goals, such as the advance of science and incremental innovations, because their publications have different citation distributions. We estimate the contribution to the progress of knowledge by studying only a limited number of the most cited papers, which are dominated by publications pursuing this progress. To field-normalize the metrics, we substitute the number of citations by the rank position of papers from one country in the global list of papers. Using synthetic series of lognormally distributed numbers, we developed the *Rk*-index, which is calculated from the global ranks of the 10 highest numbers in each series, and demonstrate its equivalence to the number of papers in top percentiles, $P_{top\ 0.1\%}$ and $P_{top\ 0.01\%}$. In real cases, the *Rk*-index is simple and easy to calculate, and evaluates the contribution to the progress of knowledge better than less stringent metrics. Although further research is needed, rank analysis of the most cited papers is a promising approach for research evaluation. It is also demonstrated that, for this purpose, domestic and collaborative papers should be studied independently.

## 1. INTRODUCTION

Developed countries invest large amounts of funds in research, and many developing countries are following the same trend. This effort by society must be compared with the output that the research system return to it, in order to assess the efficiency of the investment. On the other hand, in many research fields, countries must know their international competitive position to design their research policy.

In technological research, the product of the research system may be measured or at least estimated from patents or the economical importance of technological advances. In contrast, the product of basic scientific research is much more difficult to measure because of its intangible nature (Martin & Irvine, 1983); this uncertainty has led to a large amount of research being dedicated to its measurement. The most important conclusion of such research has been that, at a high level of aggregation, viz. countries or institutions, citation counts of scientific publications reveal their scientific success, thus supporting the use of citation-based metrics in research assessments (Aksnes et al., 2019; Waltman, 2016; Wilsdon et al., 2015).

Among the large number of citation-based indicators proposed to date, those reporting the number of publications in global top percentiles (starting the rank with the most cited paper) have received wide support (Bornmann et al., 2013) and shown the most practical mathematical properties (Brito & Rodríguez-Navarro, 2018). Furthermore, their calculation entails field normalization, which is a necessary requirement to compare different research fields because the number of citations varies extensively across them (Waltman & van Eck, 2019). Consequently, top percentile indicators are extensively used, for example, for the National Science Board (National Science Board, 2016) and the European Commission (European Commission, 2018), and have been validated against peer review (Rodríguez-Navarro & Brito, 2020; Traag & Waltman, 2019). The most widely used indicators are the number of papers among the top 10% or top 1% of the most cited papers worldwide (hereinafter denoted as $P_{top\ 10\%}$ and $P_{top\ 1\%}$ while P is the total number of papers).

However, despite such progress, common citation-based metrics do not translate into accurate indicators of scientific success. A clear contradiction between these metrics and scientific success has been found in the case of Japan, a country for which poor citation-based evaluations (Bornmann & Leydesdorff, 2013; Pendlebury, 2020; Rodríguez-Navarro & Brito, 2022a) are in flagrant contradiction to its high scientific level. For example, in the Science Reports of the European Commission for 2018, 2020, and 2022 (European Commission, 2018, 2020, 2022), the $P_{top\ 1\%}/P$ values for Japan remained almost stable during the period 2000–2018 at slightly below 0.005. These data would lead to the conclusion that Japan is a scientifically developing country despite having received numerous Nobel Prizes, that is, 17 since 2000

(https://www.nobelprize.org/ accessed 21 April 2023). This is an important contradiction that raises the question of whether currently used citation-based metrics are always reliable indicators for use in research assessment exercises.

**1.1. Contribution to the progress of science and citation-based metrics**

Although terms such as "research performance," "scientific performance," and "scientific importance" have been widely used for many years since the outset of scientometrics (e.g., Aksnes et al., 2023; Bazeley, 2010; Crespo & Simoes, 2021; Irvine & Martin, 1989; Leydesdorff, 1988; Mcalister et al., 1983; Narin & Hamilton, 1996; Taylor & Ellison, 1967), their meaning is not clear unless one defines what is to be measured. Considering that the progressive character of science makes it different from other domains of human culture (Niiniluoto, 2007), the contribution to the progress of knowledge should be one of the main objectives of research assessments. Unfortunately, this is not an easy goal to achieve because the boundaries of knowledge are pushed by very infrequent publications that are considered to be breakthroughs or landmarks (Bornmann et al., 2018). Fortunately, such publications can be recognized because they are very highly cited (e.g. Bornmann et al., 2018; Hollingsworth, 2008; Min et al., 2021; Schneider & Costas, 2017; Wuestman et al., 2020). Although not all breakthroughs are very highly cited (Hu & Rousseau, 2019), especially shortly after publication (Wang et al., 2017), and not all very highly cited papers are scientific breakthroughs (Garfield, 1973), at the country or institutional level, the positive and negative deviations cancel out and the number of very highly cited papers could be used as convenient metrics to assess the contribution to the progress of science. However, although the use of very highly cited publications for research assessment seems sound, its practical application is difficult because these publications are very infrequent and there is no statistically robust method for counting them in most countries.

The use of percentile indicators allows one to numerically describe the challenge in evaluating the contribution to the progress of knowledge. Robust statistical evaluations use the number of top 1% or top 10% most cited papers while breakthrough or landmark publications may represent only 0.02% or 0.01% of all publications (Bornmann et al., 2018; Brito & Rodríguez-Navarro, 2018). A similar conclusion is reached when considering the scientific quality of papers cited in patents: the most influential papers are in the top 0.01% most cited papers (Poege et al., 2019). In other words, major contributions to the progress of science are based on papers that may be 100 or 1000 times less frequent than the papers considered in the most commonly applied citation-based metrics. The mathematical properties of top percentile indicators may nullify this difficulty because such indicators are linked by a power-law function, and rankings based on them are independent of the percentile used (Brito & Rodríguez-Navarro, 2018; Rodríguez-Navarro & Brito, 2021b). Consequently, the expected frequency for very narrow percentiles, such as $P_{top\ 0.1\%}$ or $P_{top\ 0.01\%}$, can be calculated easily (Rodríguez-Navarro & Brito, 2019). However, deviations may hamper the general application of these mathematical properties.

## 1.2. "Perfect" research systems and synthetic series

The reasoning supporting the suggestion that citation-based metrics of normal research allows one to calculate or at least estimate the contribution to the progress of knowledge is based on the assumption that citations of scientific papers are lognormally distributed. However, although this distribution is solidly established (Golosovsky, 2021), in some fields there may be significant deviations (Waltman et al., 2012), possibly because not all researchers pursue the same objective. In technologically advanced countries at least, research also pursues incremental innovations, which are reported in papers that are unlikely to be highly cited (Rodríguez-Navarro & Brito, 2021a). This seems to be the case for Japan (Rodríguez-Navarro & Brito, 2022a), which appears to combine a potent system that addresses the progress of science alongside another pursuing incremental innovations.

Systems in which all researchers pursue the progress of knowledge may be called "perfect" or "ideal" research systems and can be mimicked by synthetic series of lognormally distributed numbers or by the aggregation of synthetic series with different $\mu$, $\sigma$, and $N$ parameters. In these systems, the numbers of papers in different top percentiles are linked by a power law (Brito & Rodríguez-Navarro, 2018), and the total number of papers and in a single top-percentile describe the whole system (Rodríguez-Navarro & Brito, 2021b). However, many real research systems are more complex. The most evident demonstration of the existence of publications that are not lognormally distributed is the large number of uncited publications (Rodríguez-Navarro & Brito, 2022a; Waltman et al., 2012). Such uncited publications can appear in research that pursues the progress of knowledge (Golosovsky & Larivière, 2021; Katchanov et al., 2023, and references therein), but also because not all research pursues the progress of knowledge. Therefore, rankings based on $P_{top\ 10\%}$ and the $P_{top\ 10\%}/P$ ratio are accurate in research systems that pursue the progress of science, but not in other cases.

**1.3. The extreme of the upper tail of citation distributions and purpose of this study**

Considering the relationship between the progress of knowledge and very highly cited papers (Section 1.1), in the search for an indicator that reveals the contribution to this progress, it is intuitive to focus the study on the papers at the end of the upper tail of the citation distribution (Glänzel, 2013), which is further supported by the finding that this allows one to detect the actual scientific level of Japan (Rodríguez-Navarro & Brito, 2022a). Such an approach has been previously applied (Rodríguez-Navarro, 2009; Rodríguez-Navarro & Narin, 2018) but faces the problem of being based on the number of citations, which varies widely across disciplines; one basic principle is that citation-based assessments should use field-normalized metrics (e.g. Bornmann, 2020; Ruiz-Castillo & Waltman, 2015; Schubert & Braun, 1996; Waltman & van Eck, 2019).

Therefore, here we investigate the application of the most cited papers to assess the contribution to the progress of science of countries and institutions by substituting ranks for the number of citations of the most cited papers. These ranks correspond to the global list of papers in a certain field of research when they are ordered according to their number of citations, from highest to lowest. Thus, all papers have two ranks; one refers to the global list, Rank 1, and the other to the local list, Rank 2. This study develops a simple and easy-to-calculate indicator based on the inverses of Rank 1; for validation, the results are compared with percentile indicators.

## 2. MATERIALS AND METHODS

Synthetic series of lognormally distributed numbers were generated as described previously (Rodríguez-Navarro & Brito, 2018). Each series simulates the scientific publications of an institution or country in a certain field, and each number in a series simulates the number of citations received by the corresponding publication. For the purposes of this study, we used 600 lognormally distributed synthetic series of different sizes ($N$), with 800, 400, and 200 numbers (we identify these series sequentially with two letters from *aa* to *xb*); the series with 280,000 numbers that result from the combination of the 600 series simulates the world publications. By combining several of the original series, we obtained series with 2000, 4000, and 8000 numbers. For the generation of the 600 series we used a single value of $\sigma$ of 1.1 (Radicchi et al., 2008; Rodríguez-Navarro & Brito, 2018; Thelwall, 2016a; Viiu, 2018) and values of $\mu$ distributed linearly from 4.0 to 2.0. This variation of $\mu$ represents the differences that exist between excellent versus poorly competitive institutions in most technological and scientific fields.

To study the publications and citations in the scientific fields of solar cells/photovoltaics and lithium batteries, we used the Web of Science. Searches were carried out making use of the Advanced Search tool in the Web of Science Core Collection, edition: Science Citation Index Expanded. The searches were limited to articles (DT=article) and to the years indicated in each case (e.g., PY=(2016-2018)). For domestic and international collaborative publications, we used the 75 most productive countries; the search was performed by separating the selected country from the others using the Boolean Operators AND or NOT (e.g., (CU=USA NOT (Canada OR Australia OR …..)) or (CU=USA AND (Canada OR Australia OR …)). For citation counting, we used a 4-year publication window and a 4-year citation window displaced five years with respect to the publication window (e.g. publications in 2014–2017 and citations in 2019–2022).

## 3. RESULTS

**3.1. Citation rank analysis in synthetic series**

Synthetic series of lognormally distributed numbers have been used to simulate "perfect" research systems and investigate the effectiveness of indicators (Brito & Rodríguez-Navarro, 2019, 2021). For the citation rank analysis, we followed the normal approach in citation-based metrics by ordering the series starting with the paper with the highest number of citations. Rank 1 corresponds to the rank of a simulated paper in the 600-aggregated series (280,000 simulated publications) and Rank 2 to the rank of this simulated paper in the individual series to which it belongs (a sample of 15 series is shown in *Supplementary material*, Table S1). The values of Rank 1 in each series describe the position of top simulated papers in the upper part of the global list. In the series simulating the most competitive countries, the values of Rank 1 are lower than those in the series simulating the less competitive countries. For example, regarding the four top simulated papers, in the series *aa*, they are in the top 0.1% of the global list (Rank 1 < 280); in the series *ej*, they are in the top 1% but below the top 0.1% (280 < Rank 1 < 2800), and in the series *un*, only three simulated papers are in the top 10% (Rank 1 < 28,000).

To assign an assessment value to each simulated paper we used the Rank 2/Rank 1 ratios of the first 10 simulated papers in each series. These ratios decrease almost linearly if a small number of sequential ratios is considered because Rank 2 and Rank 1 are linked by a power law with an exponent that does not differ much from 1 (Rodríguez-Navarro & Brito, 2018). Therefore, short segments of Rank 1 versus Rank 2 plots deviate slightly from a straight line that does not pass through the origin. To transform these ratios into a single number that operates as a representative indicator of the scientific contribution of each series, we calculated the geometric mean of the calculated Rank 2/Rank 1 ratios. We used the geometric mean, because in the series with the highest Rank 2/Rank 1 ratios, the first ratio was two- or even three-fold higher than the value of the second ratio (*Supplementary material,* Table S1 shows a sample of calculations with 15 series) and the geometric mean is more appropriate than the arithmetic mean to decrease the weight of such outliers. Furthermore, because Rank 2 and Rank 1 are linked by a power law, the geometric mean is also more appropriate than the arithmetic mean to obtain an average value of the Rank 2/Rank 1 ratios. The use of 10 simulated papers to calculate the mean was only an option considering the variability of the ratios in real cases, but other number of Rank 2/Rank 1 ratios may be used.

On the basis of the geometric mean, the indicator based on Rank 2/Rank 1 ratios could be simplified because the numerator of the geometric mean was the factorial of the number of papers considered. Thus, on the basis of 10 simulated papers, the difference between the indices based on Rank 2/Rank 1 ratios and on the inverses of Rank 1 was the tenth root of 10! (4.529).

## 3.2. Rank analyses and top percentile indicators

To confirm that the calculated geometric means of the inverses of Rank 1 were useful indicators for research assessment and not a casual occurrence, we compared them with top percentile indicators. For this purpose, we selected 99 series out of the total of 600 series, in sets of three series with the same $\mu$ parameter and $N$ equal to 800, 400, and 200; for each series we calculated the geometric means of 1/Rank 1 (10 data points) and several $P_{top\ x\%}$ values, from 15% to 0.1% (*Supplementary material*, Table S2). For this test, we obtained the $P_{top\ x\%}$ values as numbers that are not integers and can be less than 1. These non-integer $P_{top\ x\%}$ values have their origin in the lognormal distributions. In these distributions, the probability of a paper exceeding a certain number of citations can be calculated, for example, by considering the threshold of simulated citations for each $P_{top\ x\%}$. Then, the value of $P_{top\ x\%}$ can be obtained by multiplying the probability by the $N$ value of the series.

The plot of $P_{top\ 10\%}$ versus the geometric mean of 1/Rank 1 shows that the series with $N$ equal to 800, 400, and 200 follow different patterns, which collapse onto a single line

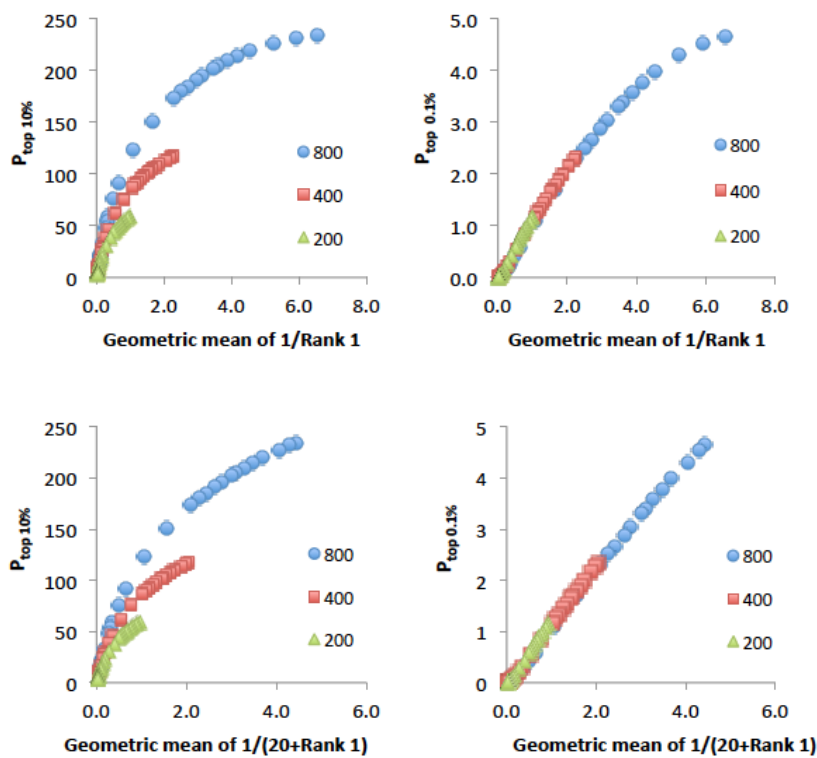

Figure 1. Plots of $P_{top\ 10\%}$ and $P_{top\ 0.1\%}$ versus the geometric means of 1/Rank 1 and 1/(20+Rank 1). Each data point correspond to the mean of the 10 highest ratios in one of the 99 synthetic series selected. Rank 1 is the rank of a simulated paper according to its number of simulated citations in the combined list of the 600 synthetic series used in this study

for $P_{top\ 0.1\%}$ (Figure 1). At low and medium values of the means, the line is straight, but it bends at high values of the means. This deviation from the straight line suggested a certain deviation of this index, which prompted us to find a correction. Because the cause probably arose from the large proportional increments that exist between the first ranks: 1 versus 2, 2 versus 3, etc., in comparison with ranks with larger numbers, the obvious correction was to add a constant number to all ranks; an empirical approach showed that addition of a constant value of 20 satisfied the requirement. The addition of 20 to Rank 1 does not change the aspect of the $P_{top\ 10\%}$ plots, but the curve in the $P_{top\ 0.1\%}$ plot disappears (Figure 1). The geometric mean of the inverses of 20 plus Rank 1 was called the "$Rk$-index" (for rank index), and we continued our study using this index.

$$Rk\text{-index} = 1000 \cdot \sqrt[10]{\prod_{i=1}^{i=10} 1/(20 + Rank\ 1i)} \qquad (1)$$

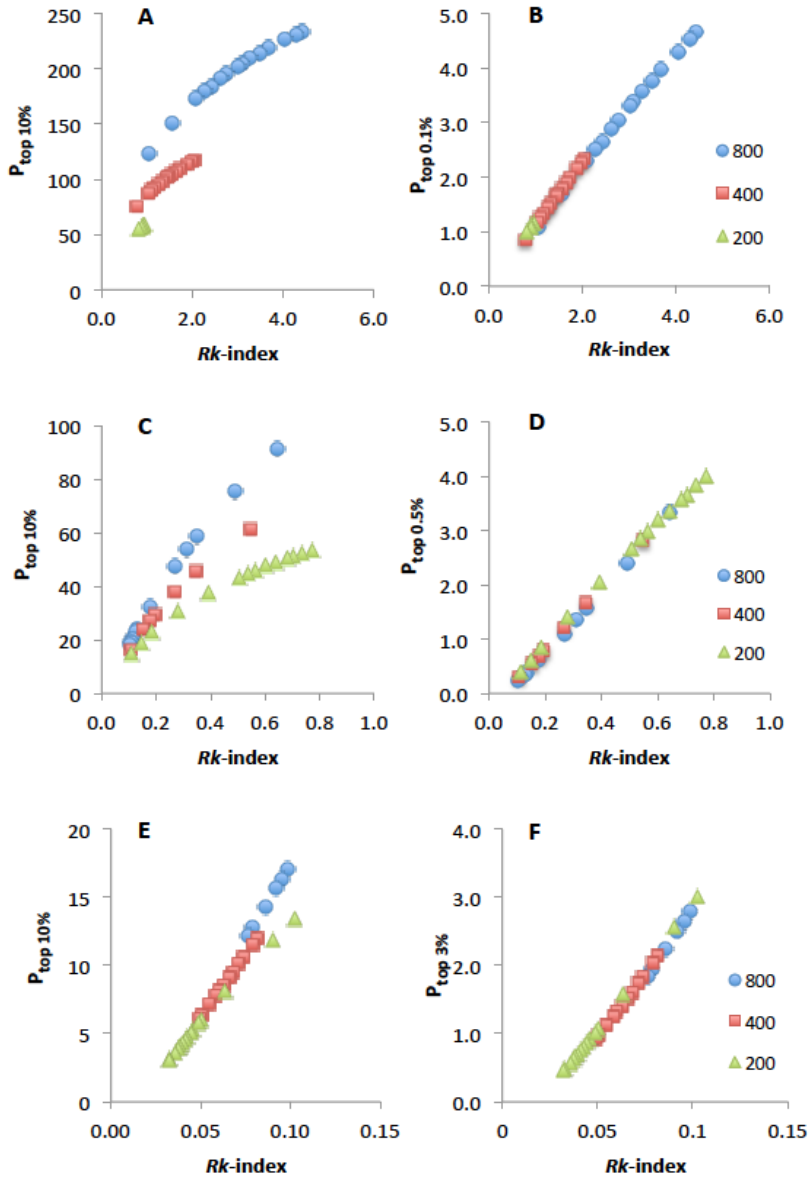

Figure 2. Plots of $P_{top\ x\%}$ versus the $Rk$-index in three levels, high (panels A and B), medium (panels C and D), and low (panels E and F). Data obtained with synthetic series as in Figure 1

Despite the apparent linearity, the ratio between the $Rk$-index and percentiles varied depending on the $\mu$ and $N$ parameters of the series. To investigate these deviations we divided the 99 series into three sets of 33 series with decreasing $Rk$-index and plotted the $Rk$-index against several $P_{top\ x\%}$. In the first set, the 800, 400, and 200 series merge for $P_{top\ 0.1\%}$; while in the second and third sets the merging occurs for $P_{top\ 0.5\%}$ and $P_{\_top\ 3\%}$, respectively (Figure 2). These results indicate that the stringency of the $Rk$-index varies; it is less stringent at lower values.

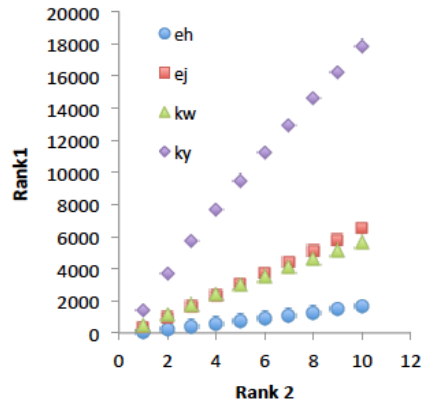

Figure 3. Effect of size and efficiency on Rank 1. Plot of Rank 1 versus Rank 2 in three synthetic series with different $\mu$ and $N$ parameters. Series *eh, ej, kw,* and *ky*: $\mu$ = 3.63, 3.63, 3.03, and 3.03; N = 800, 400, and 200; *Rk*-index = 1.56, 0.39, 0.39, and 0.12, respectively. Rank 1 is described in the legend to Figure 1; Rank 2 is the rank in the specific series

On the basis of the properties of lognormal distributions, it is obvious that the number of papers that exceed a certain threshold depends on $\mu$ ($\sigma$ is almost constant; Section 2) and $N$ parameters. To demonstrate that the *Rk*-index reveals this fact, Figure 3 shows the Rank 1 values of the 10 first simulated papers of four series selected by combining two $\mu$ values (3.63 and 3.03) and two $N$ values (800 and 200). In the series with high $\mu$ and low $N$, and low $\mu$ and high N, the Rank 1 values are almost coincident. Consistent with this coincidence, the *Rk*-indexes are very similar: 0.394 and 0.389, respectively; in the other series the *Rk*-indexes are 1.56 and 0.12 (Figure 3). In real research systems, the $\mu$ and $N$ parameters of synthetic series are equivalent to their efficiency and size, respectively. Thus, the contribution to the progress of knowledge depends on size and efficiency and the same applies to the *Rk*-index.

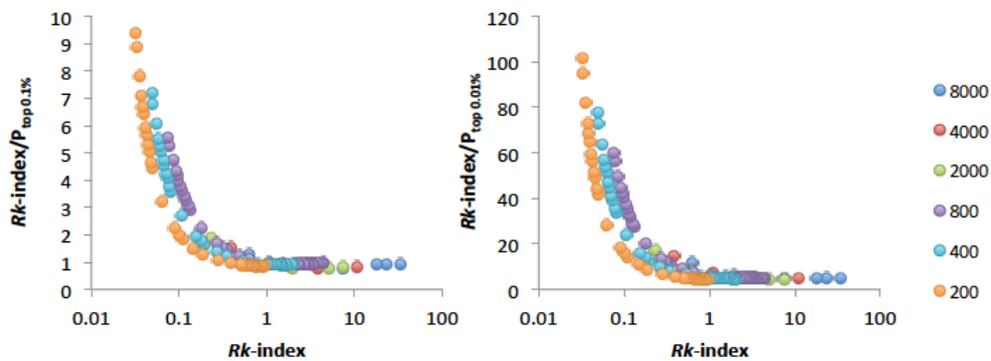

Figure 4. Plots of the ratios of the *Rk*-index with $P_{top\ 0.1\%}$ and $Pt_{op\ 0.01\%}$ versus the *Rk*-index in 115 lognormal synthetic series. The $\mu$ of the series varies from 4.00 to 2.22 and the number of simulated papers from 200 to 8000

Finally, the plots in Figure 2 raise the question of the limits of the *Rk*-index between which it is equivalent to $P_{top\ 0.1\%}$ or $P_{top\ 0.01\%}$. To answer this question, we plotted the *Rk*-index/$P_{top\ 0.1\%}$ and *Rk*-index/$P_{top\ 0.01\%}$ ratios versus the *Rk*-index (Figure 4). The plots show that between 39.5 (maximum value) and 0.5, the *Rk*-index may be taken as equivalent to $P_{top\ 0.1\%}$; between 39.5 and 1.0, the *Rk*-index can be taken as equivalent to $P_{top\ 0.01\%}$. At lower values, the *Rk*-index should be corrected before using it as equivalent to $P_{top\ 0.1\%}$ or $P_{top\ 0.01\%}$.

### 3.3. The *Rk-index* in real cases

To test the *Rk*-index in real cases, we selected two topics: solar cells/photovoltaics and lithium batteries, in which technological progress has been intense over many years. We first tested the stability of this index over three periods of four years in domestic publications (2010–2013, 2012–2015, and 2014–2017), in which the *Rk*-index does not need further calculations (Tables 1 and 2). Because the *Rk*-index reflects the competition between countries for the best ranks, in fast-evolving fields the stability of such an index based on only 10 domestic papers will be affected by the irregularity of discoveries and by the dynamics of international collaborations (these papers are in the global list but not in the domestic lists). It is worth noting that if the *Rk*-index of a country increases, it must decrease in other countries, especially if the changes occur in the most competitive countries. Therefore, we selected five countries that are leaders in these technological and scientific topics: the USA, South Korea, China, Japan, and Germany, plus a small country, Singapore, with a very active research. In both fields, the *Rk*-index shows a much higher stability than the number of publications. In both fields, between the 2010–2013 and 2014–2017 periods, the number of publications varied very significantly, in different ways across countries, but the *Rk*-index show a notable stability. Its variation was probably due to changes in the efficiency of the country.

Next, we investigated in more detail the publications of the same six countries, including in the study uncited papers ($P_0$) and $P_{top\ 10\%}$ to obtain a broader view of the relationship of these indicators. First, we studied domestic papers, but one must also consider that currently a large proportion of scientific publications are multinational collaborations (Olechnicka et al, 2019), requiring a specific treatment. The traditional method of fractional counting applied to each individual publication (Waltman & van Eck, 2015) cannot be used with the *Rk*-index because it is calculated on a whole set of 10 papers. Therefore, the correction must be applied to the calculated index by multiplying it by a correction factor considering the number of local and total addresses or authors, although the significance of this correction is not clear. In either domestic or collaborative papers, the *Rk*-index depends on the size and efficiency of the system; the size may be fractionally distributed between participating countries or institutions, but it seems unlikely that the same treatment could be applied to the efficiency. We thus studied domestic and collaborative papers independently.

Table 1. Temporal evolution of the *Rk*-index of domestic publications in six selected countries in the research field of solar cells/photovoltaics

|  | 2010-2013 | | 2012-2015 | | 2014-2017 | |
|---|---|---|---|---|---|---|
| Country | P | *Rk*-index | P | *Rk*-index | P | *Rk*-index |
| USA | 4391 | 21.04 | 5387 | 25.68 | 5505 | 25.05 |
| South Korea | 2980 | 5.14 | 4011 | 16.60 | 4208 | 17.37 |
| China | 6212 | 13.20 | 9602 | 9.54 | 12883 | 13.10 |
| Japan | 2015 | 3.74 | 2575 | 4.76 | 2751 | 4.01 |
| Germany | 1519 | 4.24 | 1796 | 2.54 | 1841 | 2.44 |
| Singapore | 406 | 2.54 | 459 | 2.51 | 365 | 1.92 |

[a] Four year citation windows displaced five years with reference to the publication window, i.e. 2015-2018, 2017-2020, and 2019-2022, respectively

Table 2. Temporal evolution of the *Rk*-index of domestic publications in six selected countries in the research field of lithium batteries

|  | 2010-2013 | | 2012-2015 | | 2014-2017 | |
|---|---|---|---|---|---|---|
| Country | P | *Rk*-index | P | *Rk*-index | P | *Rk*-index |
| USA | 1880 | 32.69 | 2775 | 33.87 | 3376 | 32.22 |
| China | 4493 | 21.65 | 8459 | 18.46 | 12577 | 20.63 |
| Germany | 417 | 5.44 | 798 | 6.21 | 1142 | 5.56 |
| South Korea | 1210 | 6.68 | 1885 | 6.07 | 2243 | 4.76 |
| Japan | 955 | 9.49 | 1177 | 6.51 | 1344 | 4.65 |
| Singapore | 281 | 7.52 | 385 | 5.84 | 317 | 3.04 |

[a] Four year citation windows displaced five years with reference to the publication window, i.e. 2015-2018, 2017-2020, and 2019-2022, respectively

Table 3. *Rk*-index and other bibliometric parameters in six selected countries in the research field of solar cells/photovoltaics[a]

|  | Domestic Publications | | | | | Internationally collaborative publications | | | | |
|---|---|---|---|---|---|---|---|---|---|---|
| Country | P | $P_0$ | $P_{top\ 10\%}$ | $P_{top10\%}/P$ | *Rk*-index | P | $P_0$ | $P_{top\ 10\%}$ | $P_{top\ 10\%}/P$ | *Rk*-index |
| USA | 5505 | 237 | 773 | 0.140 | 25.05 | 5108 | 179 | 873 | 0.171 | 19.78 |
| China | 12883 | 991 | 1118 | 0.087 | 13.10 | 4571 | 135 | 862 | 0.189 | 15.23 |
| Japan | 2751 | 335 | 163 | 0.059 | 4.01 | 1438 | 55 | 184 | 0.128 | 9.82 |
| S. Korea | 4208 | 588 | 286 | 0.068 | 17.37 | 1663 | 87 | 208 | 0.125 | 9.01 |
| Germany | 1841 | 118 | 145 | 0.079 | 2.44 | 2331 | 72 | 292 | 0.125 | 8.92 |
| Singapore | 365 | 17 | 49 | 0.134 | 1.92 | 758 | 37 | 151 | 0.199 | 4.82 |

[a] Publications in 2014-2017, citations in 2019-2022; $P_0$, uncited publications

Table 4. *Rk*-index and other bibliometric parameters of six selected countries in the research field of lithium batteries[a]

|  | Domestic publications | | | | | Internationally collaborative publications | | | | |
|---|---|---|---|---|---|---|---|---|---|---|
| Country | P | $P_0$ | $P_{top\ 10\%}$ | $P_{top\ 10\%}/P$ | *Rk*-index | P | $P_0$ | $P_{top\ 10\%}$ | $P_{top\ 10\%}/P$ | *Rk*-index |
| USA | 3376 | 44 | 580 | 0.172 | 32.22 | 2921 | 19 | 516 | 0.177 | 23.20 |
| China | 12577 | 387 | 973 | 0.077 | 20.63 | 3814 | 33 | 685 | 0.180 | 23.23 |
| Japan | 1344 | 72 | 64 | 0.048 | 4.65 | 573 | 5 | 76 | 0.133 | 4.04 |
| S. Korea | 2243 | 120 | 138 | 0.062 | 4.76 | 932 | 19 | 127 | 0.136 | 4.16 |
| Germany | 1142 | 23 | 126 | 0.110 | 5.56 | 904 | 20 | 135 | 0.149 | 13.39 |
| Singapore | 317 | 2 | 56 | 0.177 | 3.04 | 626 | 7 | 111 | 0.177 | 8.53 |

[a] Publications in 2014-2017, citations in 2019-2022; $P_0$, uncited publications

Tables 3 and 4 present the values of these *Rk*-indexes and three other indicators: P, $P_{top\ 10\%}$, and $P_0$ (the number of uncited papers). At least these four indicators for domestic and internationally collaborative papers are necessary to describe the research activity of countries and institutions. Aside from the large differences across countries, whose study lies beyond the scope of this study, on the basis of the results presented in these tables, various important conclusions can be drawn. The most important is that the *Rk*-index is absolutely independent of $P_{top\ 10\%}$, which can be expected because the path from the low stringency of $P_{top\ 10\%}$ to the high stringency of the *Rk*-index is dominated by the efficiency of the system, which can be expected to be highly variable across countries. For example, in solar cells, domestic publications, $P_{top\ 10\%}$ in South Korea is almost four times lower than in China but the *Rk*-index is higher in South Korea. Interestingly, in both technologies, $P_{top\ 10\%}$ are higher in China than in the USA, but the *Rk*-indexes are lower in China than in the USA.

Regarding the relationship between domestic and internationally collaborative papers, there is no common pattern for all countries, or even for the two considered fields in the same country. The only conclusion that can be drawn form the data presented in Tables 3 and 4 is that, when applying a hypothetic fractional reduction to the *Rk*-index, the contribution of leader countries to the progress of knowledge is dominated by domestic publications. For the two countries with the highest proportion of $P_0$ versus $P_{top\ 10\%}$, i.e., Japan and South Korea, this proportion decreases notably in international collaborations.

## 4. DISCUSSION

### 4.1. Rank-based research indicators

The *Rk*-index described here belongs to a family of research indicators that are calculated from the ranks of publications ordered by their number of citations, with the most cited first. The *h*-index (Hirsch, 2005) is the most popular among these indicators, but there is a large number of derivatives of this index (Bormann et al., 2011). The *Rk*-index belongs to this family of indicators. The important difference between the *Rk*-index and the *h*-index, along with its derivatives, is that the latter use the number of citations in conjunction with the ranks in the lists of local papers (country, institution, or researcher) to calculate the indicator. In contrast, the *Rk*-index is calculated using the global ranks, instead of citations, in addition to the local ranks. This independence from the number of citations eliminates any influence on the indicator from the significant variations in citation practices across different research fields.

**4.2. Uncertain interpretations of some citation metrics**

The inability of current indicators of research assessment to detect the scientific level of Japan (Section 1) is a warning signal about the use of these indicators. This signal deserves a careful attention because misleading research assessments can misguide research policymakers. This is the case of the false "European paradox," which has misguided the research policy of the EU for almost 25 years (Bonaccorsi, 2007; Dosi et al., 2006; Herranz & Ruiz-Castillo, 2013; Rodríguez-Navarro & Narin, 2018). Also, the use of inappropriate research indicators led to the erroneous conclusion that research policy has led to a decline of Australian research in a period when the research was actually improving (van den Besselaar et al., 2017); because of the same mistake, the research policy in Spain keeps research stuck at a intermediate level, which may even be descending (Rodríguez-Navarro, 2009, 2022). Considering the relevance of these cases, many others might exist but go unnoticed because they have not been published or even studied.

Behind this problem may lie the ambiguity of research assessments. Although there is an extensive consensus regarding the reliability of citation-based metrics in research assessments (Aksnes et al., 2019; Aksnes et al., 2023; Waltman, 2016; Wilsdon et al., 2015), albeit always in terms of correlation (Thelwall, 2016b), most research evaluations reveal a certain degree of ambiguity regarding what they measure. This problem arises because research is multifaceted (e.g. Martin, 1996; Schmoch et al., 2010; Wilsdon et al., 2015) and does not always pursue the same objective (Rodríguez-Navarro & Brito, 2021a), but most evaluations, e.g., country rankings, do not define the facet of research on which they are constructed. For example, a highly cited paper by David King (King, 2004) uses $P_{top\ 1\%}$ as the citation metric and the results are interpreted in terms of impact, high citation rate, performance, output and outcomes, and other similar features, which are unclear. In other cases, the discussion involves the term "productivity" (Daraio, 2019), whose ambiguity appears because it is not clear which type of research is evaluated. The ambiguity of the measures explains why citation metrics do not predict the number of Nobel Prizes of Japan: "Even though the number of Nobel laureates in a country and citation impact is used as indicators (proxies) for measuring the quality of research, they appear to measure different aspects of quality" (Schlagberger et al., 2016, p. 731). This conclusion raises a question about which aspects of quality are measured by Nobel Prizes and which by citation metrics.

To overcome this ambiguity, we use terms such as the "contribution to the progress of science" or "to push the boundaries of knowledge", or similar ones. The US National Science Board used the term "transformative research" (National Science Board, 2007), which is based on the same principles as the terms we use herein (Section 1.1). All these terms are coincident with the basis for awarding the Nobel Prizes. In the Nobel Prize Award Ceremony 2022, the Chairman of the Board of the Nobel Foundation said: "facing this multitude of crises and challenges, the world needs dedicated scientists who relentlessly seek the truth and push the boundaries of our knowledge"(https://www.nobelprize.org/ceremonies/opening-address-2022/, accessed on 05/1/2023). Because of the low frequency of Nobel Prizes, statistical correlations between citation metrics for the contribution to the progress of science and the number of Nobel Prizes cannot be robustly established.

**4.3. Current research indicators versus the *Rk*-index**

In contrast to many research evaluations that rank countries without mentioning a clear connection between their metrics and the contribution to the progress of knowledge, a recent research assessment by the *International Cyber Policy Centre* of the *Australian Strategic Policy Institute* (Gaida et al., 2023) explicitly addresses the evaluation of scientific advances and breakthroughs. To achieve this goal, the analyses are based on $P_{top\ 10\%}$ and the *h*-index. The *h*-index is not included in Tables 3 and 4 because its misleading use is fully documented (Brito & Rodríguez-Navarro, 2021 and references therein), but $P_{top\ 10\%}$ is included and can be used to discuss the results of the *Cyber Policy Centre*. Among the 44 technologies included in the study, "electrics batteries" and "photovoltaics" are coincident with the technologies in Tables 3 and 4. According to the study of the *Cyber Policy Centre*, China is ahead of the USA with a share of 39% versus 9% in photovoltaics and 65% versus 12% in electric batteries. In our study (Tables 3 and 4), if the comparison is based on $P_{top\ 10\%}$ in domestic papers, the results are similar to those of the *Cyber Policy Centre* results; in international collaborations, China and the USA are similar. Thus, if we based our comparison of China and the USA on $P_{top\ 10\%}$, we would reach a similar conclusion to that of the *Cyber Policy Centre*; the small differences may be introduced by the consideration of the *h*-index by the *Cyber Policy Centre*. On the contrary, if considering the *Rk*-index, for many years the USA has been ahead of China in both fields (Tables 1 and 2).

The *Rk*-index should be improved to be a perfect indicator for the contribution to the progress of science (see below), but the results obtained using the *Rk*-index are consistent and supported by the simple observation of the position of the most cited papers of countries in the global list of papers. If, in a certain field, the most cited papers in the USA are ahead of those of China in the global list of papers (smaller Rank 1 values in the USA), it can reasonably be concluded that, if the analysis is focused on breakthroughs and the contribution to the progress of science, the USA is ahead of China. In contrast, $P_{top\ 10\%}$ is not a correct metric of leadership in the contribution to the progress of science because one in 10 papers is not a breakthrough. It is worth insisting that breakthrough publications are highly infrequent (Section 1.1); therefore, a metric for breakthroughs that is based on $P_{top\ 10\%}$ requires the prior demonstration that the results based on $P_{top\ 10\%}$ are similar to those based on $P_{top\ 0.1\%}$ or $P_{top\ 0.01\%}$ (Section 1.2). This occurs for the domestic papers of some advanced countries when $P_0$ is low, but it fails in many countries. In contrast, the *Rk*-index is equivalent to $P_{top\ 0.1\%}$ and $P_{top\ 0.01\%}$, except at low values of the *Rk*-index (Figure 4).

This criticism of the use of $P_{top\ 10\%}$ is compatible with its validation against peer review in UK universities (Rodríguez-Navarro & Brito, 2020; Traag & Waltman, 2019). Firstly, most of these universities have high $P_{top\ 10\%}$/P ratios in the *Leiden Ranking* https://www.leidenranking.com/), suggesting that their research is mainly focused on the progress of knowledge, which implies that $P_{top\ x\%}$ values are linked by a power law (Brito & Rodríguez-Navarro, 2018); a similar validation might have not been possible for Nobel-level Japanese universities, for which the $P_{top\ 10\%}$/P ratios in the *Leiden Ranking* are low, at the level of developing countries. Secondly, the peer review of the UK Research Excellence Framework (REF2014, 2011) was not designed for assessment at the level of breakthroughs.

Another source of uncertainty in research assessments may arise from mixing domestic and internationally collaborative publications, even applying fractional counting, because domestic and internationally collaborative research systems may be considerably different. In some countries, the proportion of research addressing the progress of knowledge may be much higher in the latter than in the former. The high value of $P_0$ in leading research countries (Tables 3 and 4) suggests that a part of their domestic publications are addressed to push incremental innovations (next section). Regarding the contribution to the progress of science, the number of breakthroughs produced by a research system depends on its size and efficiency (Figure 3). Several methods of fractional counting seem convenient to share out correctly the size produced by the collaboration, but it is not clear that they can share out correctly the change in efficiency produced by this collaboration. A combined *Rk*-index could be obtained by summing the fractionated *Rk*-index of international collaborations to the domestic *Rk*-index, although the accuracy of the result requires investigation.

**4.4. Multiple indicators in research assessments**

A central concept in research evaluation is that a single indicator cannot reveal all types of research, which may be represented differently in different countries. Publications that pursue the advancement of knowledge are probably dominant in the citation distributions that approach to a lognormal function (Golosovsky, 2021), where the number of uncited papers is low. In contrast, publications that originate from research that pushes incremental innovations are less basic for further research, implying that many of them will not be cited, while others will receive a small number of citations (Rodríguez-Navarro & Brito, 2022a, 2022b). The number of uncited papers (Golosovsky & Larivière, 2021; Katchanov et al., 2023; and references therein) has been explained by mechanisms that do not imply two populations of papers corresponding to two types of research. The results of this study show that, in the same field, some advanced countries have much more uncited papers than others, and that the same country can show very different numbers in two different fields. For example, in Japan, the proportion of uncited domestic papers is higher in solar cells than in lithium batteries. In less advanced countries, a large number of uncited papers can also appear when there is a population of researchers with very low competitiveness (Rodríguez-Navarro, 2009). Therefore, further research in this field should also include the number of patents.

It is also worth noting that the concept of uncited papers ($P_0$) is temporal because many papers that are uncited in a three-year citation window will have one citation in a five-year citation window. $P_0$ is only an indicator of a population of papers whose citation dynamics (Golosovsky & Larivière, 2021) are different from the population of papers whose citation distribution is lognormal (Golosovsky, 2021). Thus, for evaluative purposes, the importance of $P_0$ is comparative between countries, being indicative of a population of papers that differ from those that pursue the progress of knowledge and that might be a significant proportion of all papers. $P_0$ should not be omitted from research evaluations of scientific fields of technological importance, but it must be remembered that it is only an indicator of the presence of a different population of papers. Because the number of uncited papers is temporarily variable, the exclusion of uncited papers (Thelwall, 2016a) is not a solution in the process of research evaluation. For example, Japan is a Nobel Prize-winning country in lithium batteries (Akira Yoshino, Chemistry, 2019) with high $P_0$. However, its very low $P_{top\ 10\%}$ with reference to P (Table 4) suggests that the population of papers that does not pursue the progress of science may be much larger than $P_0$.

The *Rk*-index avoids this complexity. The identification of the publications that pursue the progress of knowledge is not currently possible, but the rank analysis proposed herein is focused on them, avoiding their identification by considering the 10 most cited papers. On this basis, this study has two aims: to place the contribution to the progress of science as a useful indicator in the research assessment of countries and institutions, and the use of the rank analysis of the most cited papers as a method to estimate this contribution. This analysis is different from percentile metrics because, in those metrics all papers exceeding a certain citation threshold have the same value. Meanwhile, in citation rank analysis, each paper has its own value, which decreases with the increase of the rank number. This also occurs in other citation metrics that are field normalized (Waltman & van Eck, 2019), and specifically in the Relative Citation Rate (Hutchins et al., 2016). The major advantage of rank analysis over these other methods is that the mathematical link of the ranks (Rodríguez-Navarro & Brito, 2018) allows many different mathematical treatments of the data. The *Rk*-index may be the simplest and easiest treatment. It only requires identifying the ranks of the 10 most highly cited local papers in the citation-ordered global list of papers and calculating the geometric mean of these ranks after the addition of 20. This simplicity entails a few deviations from an ideal index. In low-level countries, none of the 10 most cited papers is highly cited, which implies that the *Rk*-index is not linearly related to the contribution to the progress of knowledge (Figure 4). The *Rk*-index is excellent to compare systems that are similar in their contribution to the progress of knowledge, but a correction is necessary to compare very dissimilar systems. Corrections of the *Rk*-index for low competitive countries or new approaches in citation rank analysis open new avenues of research in research evaluations.

Tables 3 and 4 describe the research systems in two scientific fields, and the same could be done in many other scientific or technological fields. The abovementioned report of the *International Cyber Policy Centre* studies 44 research fields; the same fields could be studied using the four indicators presented in Tables 3 and 4, or the study could be extended to more fields. Eventually, a general *Rk*-index could be obtained as a mean of all these *Rk*-indexes, expecting that this mean will provide a general idea of the capacity of each country to contribute to the progress of knowledge. Although, after solving the technical problems of publications that are common in two or more research fields, this approach may be correct in large countries because they investigate in all or most technologies, but the same reasoning may not apply to small countries.

A different approach might be to investigate general fields, namely chemistry or mathematics, although they contain many subfields with different average numbers of citations per publication. Assuming that the most competitive countries engage in research related to the most important topics for the progress of science, the *Rk*-index could be a reasonable indicator of their contribution to the progress of knowledge in the general fields analyzed.

**4.5. Lessons from two research fields**

The purpose of this study was only theoretical, but the results for solar cells and lithium batteries (Tables 3 and 4) allow some conclusions to be reached regarding the advantages of using the rank analysis of the most highly cited papers. In solar cells, the highly probable remarkable contribution of domestic papers from South Korea cannot be predicted from commonly used indicators. It is obvious that $P_{top\ 10\%}$ does not provide any information about the $Rk$-index, and it seems clear that $P_{top\ 10\%}/P$ may be highly misleading when describing the efficiency of research systems.

A second interesting finding is that, in leading countries, the contribution of domestic papers to the total of very highly cited papers is higher than that of internationally collaborative papers. In these countries, the $Rk$-index may be higher in internationally collaborative papers, but after applying a fractional reduction, the dominance of domestic papers seems to be the norm in leading research countries.

The number of uncited papers is another long-forgotten datum, which seems to be an indicator of a population of papers whose citation dynamics differ from that of papers that are contributing to pushing the boundaries of knowledge. The latter but not the former may be very highly cited. A parallel patent analysis could reveal whether the poorly cited population of papers pursues the push towards incremental innovations.

All these results strongly suggest that comprehensive research evaluations of countries should include more indicators than the number of papers in a single percentile. The $Rk$-index or other indicator based on the rank analysis of the most cited papers appear to be convenient indicators to include.

**Acknowledgments**

We acknowledge a fruitful discussion of one of us (A R-N) with Francis Narin regarding the use of top cited papers for research evaluation.

**Author contributions**

Alonso Rodríguez-Navarro: Conceptualization, Data curation—citation data. Formal analysis. Visualization. Writing—original draft. Writing—review & editing. Ricardo Brito: Data curation—synthetic data. Formal analysis. Funding acquisition. Visualization. Writing—review & editing.

**Competing interest**

**Funding information**

This work was supported by the Spanish Ministerio de Economía y Competitividad, Grant Number PID2020-113455GB-100.

**Supplementary materials**

Supplementary material associated with this article can be found, in the online version, at xxxxxxxxx

Niiniluoto, I. (2007). Scientific Progress. In E. N. Zalta (Ed.), *The Stanford Encyclopedia of Philosophy*. The Metaphysics Reserach Lab, Stanford University.

Olechnicka, A., Ploszaj, A., & Janowicz, D. (2019). *The Geographyof Scientific Collaborations*. Routledge , Oxford and Ney York.

Pendlebury, D. A. (2020). When the data don't mean what they say: Japan's comparative underperformance in citation impact. In C. Daraio & W. Glanzel (Eds.), *Evaluative Informetrics: The Art of Metrics-based Research Assessment*. Spriger.

Poege, F., Harhoff, D., Gaessler, F., & Baruffaldi, S. (2019). Science quality and the value of inventions. *Science Advances*, *5*, eaay7323.

Radicchi, F., Fortunato, S., & Castellano, C. (2008). Universality of citation distributions: toward an objective measure of scientific impact. *Proc. Natl. Acad. Sci. USA*, *105*, 17268-17272.

Assessment framework and guidance on submissions, (2011). https://http://www.ref.ac.uk/2014/pubs/2011-02/ accessed on October 2018

Rodríguez-Navarro, A. (2009). Sound research, unimportant discoveries: Research, universities, and formal evaluation of research in Spain. *Journal of the American Society for information Science and Technology*, *60*, 1845-1858.

Rodríguez-Navarro, A. (2022). *Cómo medir el éxito científico. Los errores de España*. Aula Magna.

Rodríguez-Navarro, A., & Brito, R. (2018). Double rank analysis for research assessment. *Journal of Informetrics*, *12*, 31-41.

Rodríguez-Navarro, A., & Brito, R. (2019). Probability and expected frequency of breakthroughs – basis and use of a robust method of research assessment. *Scientometrics*, *119*, 213-235.

Rodríguez-Navarro, A., & Brito, R. (2020). Like-for-like bibliometric substitutes for peer review: advantages and limits of indicators calculated from the ep index. *Research Evaluation*, *29*, 215-230.

Rodríguez-Navarro, A., & Brito, R. (2021a). The link between countries' economic and scientific wealth has a complex dependence on technological activity and research policy. *Scientometrics*, *127*, 2871-2896.
28

30

Traag, V. A., & Waltman, L. (2019). Systematic analysis of agreement between metrics and peer review in the UK REF. *Palgrave Communications*, *5*, 29. https://doi.org/ https://doi.org/10.1057/s41599-019-0233-x

van den Besselaar, P., Heyman, U., & Sandström, U. (2017). Perverse effects of output-based research funding? Butler's Australian case revisited. *Journal of Informetrics*, *11*, 905-918.

Viiu, G.-A. (2018). The lognormal distribution explains the remarkable pattern documented by characteristic scores and scales in scientometrics. *Journal of Informetrics*, *12*, 401-415.

Waltman, L. (2016). A review of the literature on citation impact indicators. *Journal of Informetrics*, *10*, 365-391.

Waltman, L., & van Eck, N. J. (2015). Field-normalized citation impact indicators and the choice of an appropriate counting mehod. *Journal of Informetrics*, *9*, 872-894.

Waltman, L., & van Eck, N. J. (2019). Field Normaliation of Scientometric Indicators. In W. Glänzel, H. F. Moed, U. Schmoch, & M. Thelwall (Eds.), *Springer Handbook of Science and Technology Indicators*. Springer Nature.

Waltman, L., van Eck, N. J., & van-Raan, A. F. J. (2012). Universality of citation distributions revisited. *Journal of the American Society for information Science*, *63*, 72-77.

Wang, J., Veugelers, R., & Stephan, P. (2017). Bias against novelty in science: A cautionary tale for users of bibliometric indicators. *Research Policy*, *46*, 1416-1436.

Wilsdon, J., Allen, L., Belfiore, E., Campbell, P., Curry, S., Hill, S., Jones, R., Kerridge, S., Thelwall, M., Tinkler, J., Viney, I., Wouters, P., Hill, J., & Johnson, B. (2015). *The metric tide: Report of the independent review of the role of metrics in research assessment and management*. https://doi.org/10.13140/RG.2.1.4929.1363

Wuestman, M., Hoekman, J., & Frenken, K. (2020). A topology of scientific breakthroughs. *Quantitative Science Studies*, *1*, 1203-1222.